# One pocket to activate them all: Efforts on understanding the modulator pocket in K2P channels

Edward Mendez-Otalvaro[a], Wojciech Kopec[a,b*], Marcus Schewe[c*], and Bert L. de Groot[a*]

[a]*Computational Biomolecular Dynamics Group, Max Planck Institute for Multidisciplinary Sciences, Göttingen, Germany;* [b]*Department of Chemistry, Queen Mary University of London, London, United Kingdom;* [c]*Institute of Physiology, Kiel University, Kiel, Germany.*

*\*To whom correspondence may be addressed. Email: w.kopec@qmul.ac.uk, m.schewe@physiologie.uni-kiel.de and bgroot@gwdg.de.*

# One pocket to activate them all: Efforts on understanding the modulator pocket in K2P channels

The modulator pocket is a cryptic site discovered in the TREK1 K2P channel that accommodates agonists capable of increasing the channel's activity. Since its discovery, equivalent sites in other K2P channels have been shown to bind various ligands, both endogenous and exogenous. In this review, we attempt to elucidate how the modulator pocket contributes to K2P channel activation. To this end, we first describe the gating mechanisms reported in the literature and rationalize their modes of action. We then highlight previous experimental and computational evidence for agonists that bind to the modulator pocket, together with mutations at this site that affect gating. Finally, we elaborate how the activation signal arising from the modulator pocket is transduced to the gates in K2P channels. In doing so, we outline a potential common modulator pocket architecture across K2P channels: a largely amphipathic structure -consistent with the expected properties of a pocket exposed at the interface between a hydrophobic membrane and the aqueous solvent- but still with some important channel-sequence-variations. This architecture and its key differences can be leveraged for the design of new selective and potent modulators.

**Introduction**

Two-pore-domain potassium (K2P) channels are a family of potassium ($K^+$) channels found in eukaryotes, responsible for leak currents that regulate the resting membrane potential [1]. Also known as leak-current channels, these were first discovered in 1995 in baker's yeast [2]. K2P channels exhibit an atypical topology compared to other $K^+$ channel families: they consist of two monomers, each with four transmembrane helices (M1-M4), two pore domains (P1 and P2), two selectivity filter (SF) strands (SF1 and SF2), and an extracellular M1-P1 loop, with two loops assembling into the characteristic CAP structure of the K2P channels. Upon assembly, the dimer forms a pseudo-tetramer [3], which can exist as either a homodimer or heterodimer, thus enhancing channel-expression variability [4]. According to multiple structural studies, some K2P channels can display their M4 helix rotating and shifting upward (toward) the M2 of the opposing monomer (called the up-state), whereas the down-state positions M4 nearly 45° with respect to the membrane. The up-state shows higher conductance compared to the down-state [5–7] while channel activation can also emerge from the down-state [8].

The K2P channel family in humans consists of 15 different channel types [9], which are distributed across various tissues, including the heart [10,11] and the central nervous system [12], where they are found in cells such as astrocytes, neurons, and microglia [13]. These channels are sensitive to various external stimuli, including pressure, temperature, and changes in intracellular pH ($pH_i$) and extracellular pH ($pH_e$) [14–18], suggesting a complex mechanism that transduces external signals into changes in the channel's activity.

The functionality of K2P channels critically contributes to the resting membrane potential as well as the shape and frequency of action potentials. Since the resting membrane potential largely determines whether a neuron fires an action potential, K2P channels represent attractive drug targets, and their modulation holds promise as an attractive strategy for treating disorders of cellular excitability [19–22]. Several developmental, metabolic, and neurological conditions have been linked to mutations in K2P channels [23]. For example, mutations that reduce outward currents in the TASK3 K2P channel cause a syndrome characterised by impaired growth, metabolic dysregulation, developmental delays, and cognitive deficits [24]. Mutations near the SF of TRAAK are associated with facial dysmorphism, hypertrichosis, epilepsy, intellectual disability, and gingival overgrowth [25]. Similarly, a point mutation within SF2 in TREK1 induces unusual sodium permeability and hypersensitivity to membrane stretch in the channel [26]; while certain loop mutations in TRESK have been implicated in migraine [27]. In contrast, gain-of-function (g-o-f) mutations in TALK1 and TASK1 have been associated with metabolic or developmental conditions, including maturity-onset diabetes of the young [28], and developmental delay with sleep apnea [29], respectively. Together, these findings suggest the therapeutic potential of K2P channels and have motivated efforts to develop targeted modulators.

**Figure 1.** *Different sequences in the selectivity filter, but the same family of K⁺ channels.* **Middle panel:** Snapshot of a K2P channel embedded in a phospholipid membrane, the inset highlights the SF and pore region. **Surrounding upper panels**: Cylindrical cartoon representations of the SF1s in different K2P channels according to structures deposited in the PDB. The bolded name indicates the channel family (gene), while the name in parentheses refers to the specific channel corresponding to the snapshot shown. The presence of K⁺ were preserved as they appear in the structural file. In magenta key residues are highlighted. The cartoon representation was colored based on the Eisenberg hydrophobic scale [30], which is found as a color bar at the bottom of the panels (the whither, the more hydrophobic). Residues ARG, LYS and HIS were colored blue, while residues ASP and GLU were colored red, in order to highlight potential electrostatic interaction sites. The one-letter code corresponds to sequence differences present in the P1 helix and the SF1. The dashed orange-edged snapshot in the lower-left panel represents the aligned SF2 for all channels shown, highlighting the conservation of aspartate in the SF2, a hydrogen bond donor/acceptor (tyrosine) in the P1 helix, and the absence of a hydrogen bond donor/acceptor residue behind the SF2 (valine/isoleucine). This contrasts with the variability shown by the channels in the equivalent SF1 region. **Lower panel:** Comparison of the sequences of the K2P channels shown in the upper panel. The magenta arrows highlight the one-letter-code residues shown in the upper panel**.** The sequencing in SF1 (upper sequence-comparison) is variable across the different channels, including the residue located in the P1 helix that interacts with the residue in the upper part of the SF1. In contrast, the SF2 (lower sequence-comparsion) shows conserved features: A hydrogen bond donor/acceptor on the P2 helix (TYR hydroxyl) which interacts with the residue on top of the SF2 (ASP carboxylate). A hydrophobic residue (ILE/VAL) is also conserved behind the SF2, as opposed to a hydrogen bond donor/acceptor residue (THR hydroxyl) behind the SF1. A green star highlights the hydrophobic patch located beneath the residue equivalent to T141 in TREK1 -a common subregion targeted by multiple endogenous and exogenous ligands across K2P channels. The same green start is also displayed in the structural models shown in the upper panel.

## K2P channels display a gating mechanism involving the SF, akin to C-type inactivation

A distinctive feature of K2P channels is the variation in their SF sequence compared to the signature sequence (TxGxG), which is typically followed by an aspartate (ASP, D). K2P SF sequence changes are mainly located in the second and fourth positions, including substitutions such as x/V→I and x/Y→F. The ASP-flanking sequence is also highly variable among different K2P channels, with changes such as D→H (TASK1), D→N (TREK1), D→Y (TWIK2), and D→M (THIK1), for instance [31–34]. Additionally, since K2P channels form pseudo-tetramers, the four strands that make up the SF can be sequentially asymmetric within the same chain [3,35]. For this reason, it is common to divide the SF into SF1 and SF2 to distinguish between the different strands that contribute to the overall filter architecture (e.g., in TREK1, SF1 reads TIGFGN, and SF2 TIGFGD, with both strands present in chains A and B). This sequence variability across K2P channels likely affects gating, ion conduction, and selectivity, as these properties depend on the stability and composition of the SF, which, in turn, is influenced by the chemical properties of the amino acids that constitute the filter (Figure 1) [36–38].

One of the primary gating mechanisms in K2P channels involves the SF itself: distorted, asymmetric SF geometries that deviate from the canonical conducting structure hinder ion permeation through the central pathway, ultimately regulating channel conductance. This mechanism was described by *Lolicato, Natale et al.* [39] for TREK1, based on crystallographic studies under different $K^+$ concentrations [39]. This inactivation process of the filter upon inward-flux was previously suggested in the mechanism of voltage sensing for K2P channels by *Schewe et al.* [40], and more recently, further solved at atomistic level through electrophysiology experiments and molecular dynamics (MD) simulations [41]. Similarly unfavorable geometries for $K^+$ coordination in S0 and S1 binding sites were reported for TASK2 under low $pH_e$, involving dilation at the top of the SF due to rearrangements of residues in the CAP domain and M4 helix [42]. Another K2P channel, TASK3, exhibits a C-type inactivated state upon acidification and rearrangement of its pH sensor, H98. This state involves dilation of the S0 and S1 binding sites, as well as hydrophobic blockage caused by the side chains of tyrosine and phenylalanine residues that form the S1 site [43]. Similar to TASK3, rearrangements in the histidine residues following the S0 site in TASK1 have also been reported, affecting the structure of the top of the filter [31]. Comparable dilations in S0-S1 have been observed in another non-selective non-K2P channel, the *Bacillus cereus* NaK channel [44]. These structural dilations are notably similar to recent Cryo-EM structures of *Shaker* [45,46], Kv1.2 Chimera-3m [47], and Kv1.3 [48], which are considered C-type-inactivated states, in contrast to other inactivated states involving rather a

*collapse/pinching* in the middle of the filter, as seen in KcsA [49], and as suggested by MD simulations on a pore model of *Shaker* [50].

The relationship between these various non-canonical filter geometries in K2P channels, their filter sequences, and their biological function remains an underexplored research area. This topic could improve our understanding of how inactivation mechanisms vary across K2P channel subfamilies.

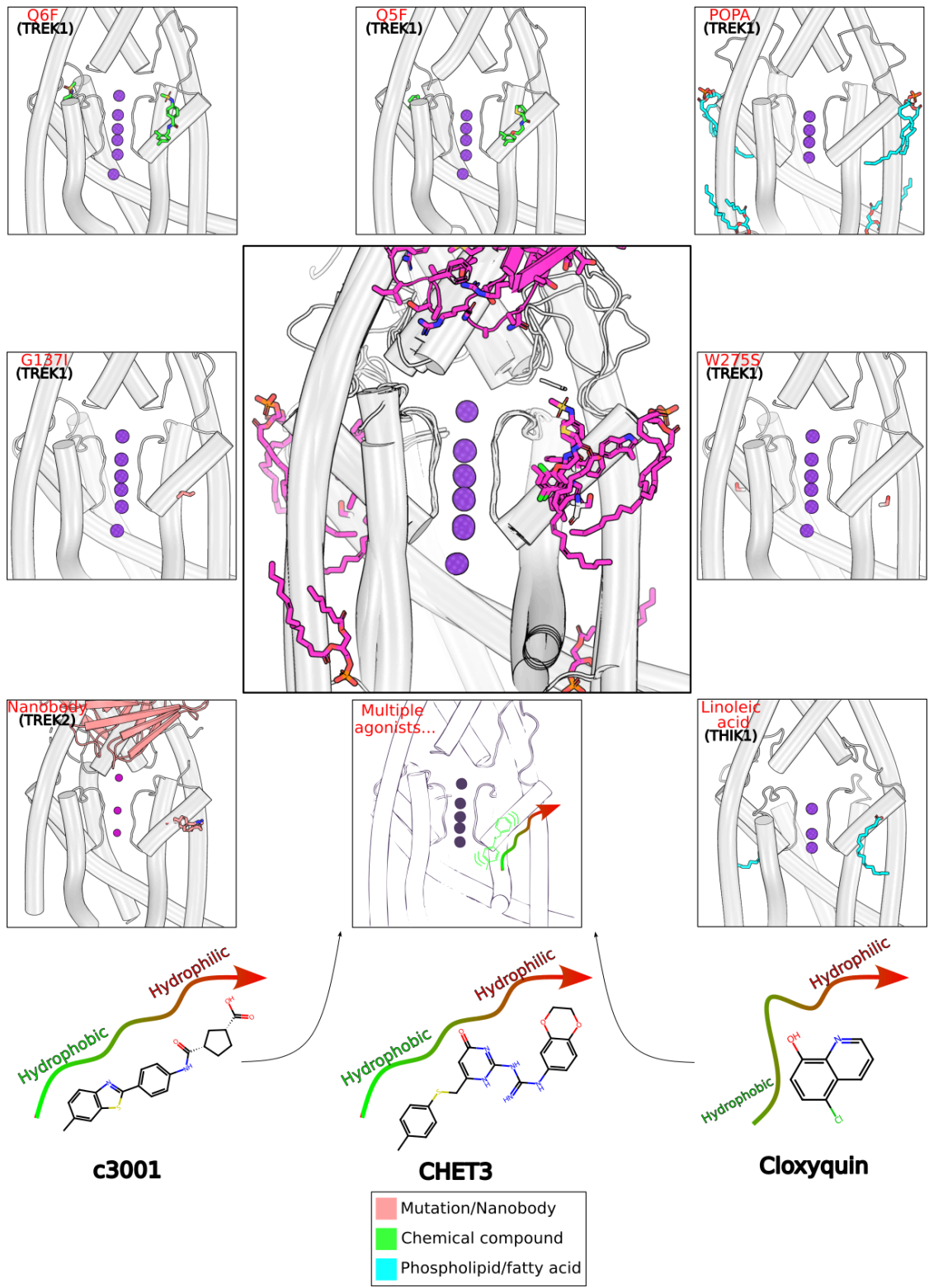

**Figure 2.** *Multiple experimental evidence of activation in K2P channels converge in the modulator pocket.* **Middle panel.** Representative snapshots of solved structures for multiple K2P channels where hydrophobic motifs in the lower modulator pocket, and more polar motifs in the upper part, activate the channel. The binder/mutation are highlighted according to the lower legend. **Surrounding panels**. Breakdown of each of the solved structures shown in the middle panel. Binder or mutation names are shown in red, while the corresponding K2P channel is indicated in black within brackets. The lower-central panel, marked with a color-gradient arrow, just indicates that the molecules listed below

bind close to the modulator pocket, but there are no experimentally-solved structures associated with them. W275S and G137I mutants (right and left surrounding-panels) were generated using PyMOL based on the TREK1 (PDB: 6CQ6) WT model. For details about each evidence please refer to the main text.

**K2P channels can be directly activated through a pocket adjacent to the SF: "The modulator pocket"**

Since one of the main gating mechanisms in K2P channels involves the filter structure, it makes sense to focus efforts on stabilizing its conducting geometry. *Lolicato, Arrigoni, et al.* [32], through crystallographic and electrophysiological work, identified a cryptic pocket -i.e., a pocket that is formed in the holo structure, but not in the apo one- adjacent to the SF1 of TREK1. This pocket is composed of P1 and M4, and they named this region the "modulator pocket" [32]. Following this important discovery, subsequent findings have demonstrated that this pocket is crucial for directly activating TREK1 and potentially other K2P channels. Various alterations in this pocket can directly influence its impact on channel activity (Figure 2).

*Mutations*

Mutations in the P1-M4 interface have been shown to activate the TREK1 channel: G137I and W275S mutants increase current amplitudes in electrophysiology measurements, respectively, and it is proposed that they are key in the communication between the C-terminal domain of the channel and the C-type inactivation mechanism in the filter of TREK channels [51–53]. Equivalent g-o-f mutations (G124I and W262S) have also been engineered in the TRAAK channel, which structural studies and whole-cell patch-clamp recordings have verified [5], although the TRAAK channel activation is based on crystal structures of the g-o-f mutations that were found to be in the down state [5]. Regardless, structurally both mutations might generate changes in the chemical environment surrounding the modulator pocket: G137I inserts a hydrophobic and bulkier motif into the pocket, whereas W275S considerably decreases the steric hindrance of the side chain in the pocket.

*Lipids and fatty acids*

The effect of hydrophobic biomolecules in the modulator pocket has been documented for TREK1 by *Schmidpeter et al.* [54], where Cryo-EM densities show that the anionic lipid 1-palmitoyl-2-oleoyl phosphatidic acid (POPA) inserts through its tail into the modulator pocket, pushing the W275 side chain towards the filter. Functional fluorescence quenching studies indicate that this lipid indeed activates the channel [54]. Similarly, the first THIK1 channel structures were independently resolved by two groups at the same time, both discovering a polyunsaturated fatty acid (modeled as a linoleic acid) that is inserted into the equivalent modulator pocket for THIK1, but in a conformation that covers mostly the lower part of the site, also demonstrating activation of the channel in electrophysiology measurements [34,55]. This evidence indicates that the modulator pocket is a favorable site for the binding of hydrophobic molecules and suggests that this physicochemical property is key to the activation of the filter through this binding event.

*Nanobodies*

*Rödström, Cloake, et al.* [56] demonstrated by Cryo-EM and whole-cell electrophysiology measurements that the binding of a nanobody (named Nb-Activator-76) activates TREK2 [56]. In detail, the nanobody penetrates the modulator pocket from the top, interacting with residues E128 and Q106 of M2, as well as N292, located in the extracellular loop connecting SF2 to M4. Notably, the nanobody pushes residue W306 (equivalent to W275 in TREK1) into the pocket while displacing F164 (equivalent to F134 in TREK1) outward. This movement packs the phenyl and indole rings of both residues in the center of the modulator pocket. This once again demonstrates that hydrophobic or bulky residues play a crucial role in the activation of TREK channels.

*Small molecules*

As mentioned earlier, the discovery of the modulator pocket started with the resolution of crystallographic models of two small molecules (ML335 and ML402) bound to TREK1, obtained by *Lolicato, Arrigoni et al* [32]. Both molecules present upper and lower aromatic rings, linked by a moiety showing a peptidic-like bond. The molecules are non-covalently bound into the pocket in an L-type conformation, whereby the upper ring interacts with K271 in the upper pocket; and with W275 and F134 with the lower ring. The compounds bind with a 2:1 stoichiometry (since the channel has two modulator pockets, one in each monomer). In electrophysiology measurements of the crystal constructs, ML335 shows an $EC_{50}$ of 10.5 ± 2.7 µM, and ML402 an $EC_{50}$ of 14.9 ± 1.6 µM, opening a door for potential modifications of these molecules to increase specifically their apparent binding affinity for TREK1, since, at least ML335, exhibits already selectivity towards TREK channels compared to TRAAK, due to the abolition of a π-cation interaction between K271 and the upper aromatic ring [32]. Further studies by *Deal et al.* [57] introduced activators covalently bound to TREK1 via adduct-formation between an acrylamide-ML335-derivative ligand (termed ML336), and the hydroxyl group of the S131 residue located on P1. Although the formation of the covalent bond was confirmed under crystallographic conditions, the ML336-S131 adduct does not form under functional study conditions. However, a derivative of the ML335 molecule with a maleimide substitution (termed CAT335) can activate the mutant TREK1 S131C channel even after washout conditions, demonstrating the formation of an adduct with the thiol group of the S131C residue, and allowing to lock

TREK1 in an active state [57]. Importantly, they found that the 2:1 stoichiometry is not strictly necessary to activate the channel, since it is sufficient to have one molecule bound to either one of the two pockets in TREK1 [57].

Interestingly, other authors have also found activators that putatively bind to this pocket. For instance, *Qiu et al.* [58] employing whole-cell patch-clamp and single-channel measurements from outside-out patches; identified an activator (C3001a) that selectively activates TREK1 ($EC_{50}$ = 12.81 μM) and TREK2 ($EC_{50}$ = 11.31 μM), while showing lower activity toward TRAAK and no activation for TASK3, TASK1, TRESK, or THIK1 [58]. Further, the authors characterized the binding mode of C3001a in the modulator pocket through a combination of mutagenesis, dose-response curves, and molecular docking. According to their results, the ligand adopts a conformation similar to that of ML335 and ML402, inserting its methylbenzothiazole ring at the bottom of the pocket. Moreover, the molecule interacts with residues Y270 and Y266 at the top of the pocket via its negatively-charged cyclopentane-carboxylate group. Importantly, the residue Y270 has been identified as crucial for translating signals such as temperature and pressure in TREK1 channel by stabilization of the "M3 glutamate network" as termed by *Lolicato et al.* [39].

Some modulators bind to a different site close to the modulator pocket, named the fenestration site. This fenestration is a "gap" formed by M2 and M4 and can be occupied by the inhibitor norfluoxetine (NFx) [6,8]. An further example is the molecule PD-118057, which is an activator in TREK channels that binds into the fenestration site, according to *Schewe et al.* [59]. More recently, this same molecule was characterized by *Liu et al.* [60] using MD simulations and functional assays. They propose that the molecule interacts with the S4-forming threonines adjacent to T141 via its carboxylate group, while its substituted phenyl ring engages with hydrophobic side chains located beneath T141 in a more dynamical fashion [60]. This molecule activates the channel even in the presence of ML335, suggesting that it interacts in a manner similar to prior reported negatively-charged activators (NCAs), which enhance $K^+$ ion occupancy along the permeation pathway [59]. These results indicate that a coupling between the modulator pocket and the fenestration site is possible. We will expand about this potential synergistic effect between binding sites later.

Other K2P channels also exhibit activation upon binding of molecules within the modulator pocket region, particularly in the lower pocket, close to the base of the SF. For example, in a study combining mutagenesis, electrophysiological assays, and MD simulations, *Schreiber et al.* [61] demonstrated that cloxyquin activates TRESK ($EC_{50}$ = 64.00 ± 11 μM) by binding close to the M2-M4 interface -a region that also partially encompasses the modulator pocket, as was previously shown for TREK1 [62]. Notably, the T322A mutation in TRESK -corresponding to TREK1 T141 residue- results in reduced ion permeation, an effect that the activator is able to compensate for, thereby restoring ion flow. The molecule engages in hydrophobic interactions between its quinoline ring and a hydrophobic patch on M4, located just below the equivalent of T141 residue in TREK1, and also forms electrostatic interactions between a hydroxyl group and a nearby aspartate on M4 [61].

Likewise, *Liao et al.* [63], using virtual screening and a homology model, identified a molecule, named CHET3, that is an activator for TASK3 channels ($EC_{50}$ = 1.4 ± 0.2 μM). Through mutagenesis, functional assays, and MD simulations, they mapped the binding site of this activator close to the lower modulator pocket. CHET3 forms hydrogen bonds via its hydroxyl groups with threonines close to the filter and engages in hydrophobic interactions through its aromatic rings with hydrophobic residues on M4 [63].

All of the above reports point to the fact that the modulator pocket is a general feature of K2P channels, but at the same time, it has its unique characteristics within each subfamily, which opens the door to selective modulators' design.

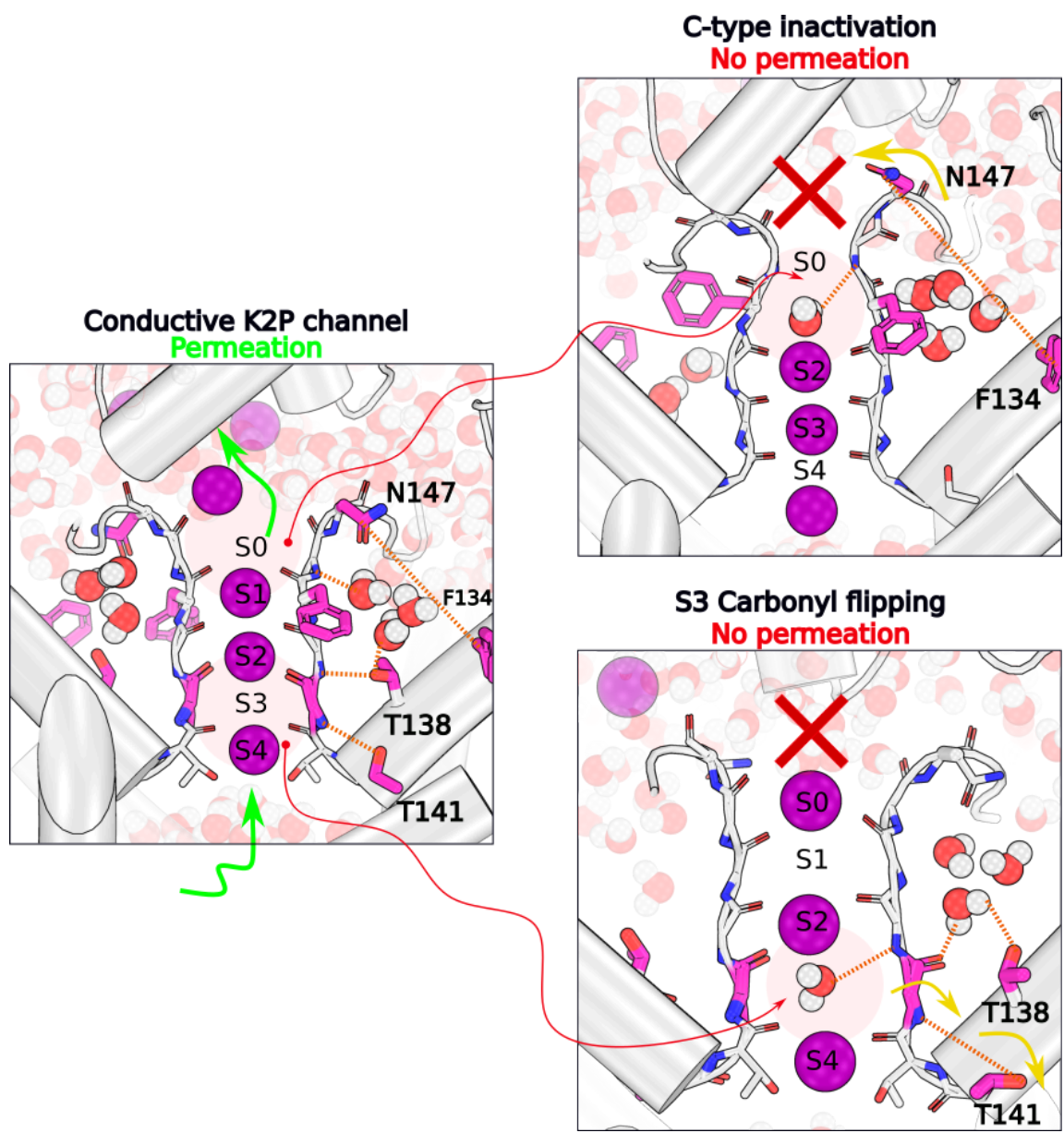

**Figure 3.** *Rationalizing the modulator pocket mechanism.* Two potential mechanisms within the SF1 may give rise to non-permeation events in K2P channels: **i) C-type inactivation**, which involves the S1 and S0 binding sites, as well as the residue flanking S0 (ASN in this example). This region is located at the top of the SF and includes interactions between the residue flanking S0 and another residue located on

P1, positioned at the top-rear of the SF1 (phenylalanine in this example). Distortions in this region -driven by the dynamics of S1, S0, and the ASN- can disrupt the permeation pathway, thereby hindering ion conduction. **ii) Carbonyl flipping at S3**, which involves the S3 binding site and a stabilizing hydrogen-bond network. This network includes a donor/acceptor residue at the base of the filter (THR in this example) and another residue located on P1 at the back-bottom of the SF1 (THR in this case). When this network is perturbed -such as by the presence of a water molecule in S3 and another just behind it- carbonyl flipping occurs, disrupting ion permeation. The modulator pocket includes residues directly involved in these two mechanisms, which may explain how binding of activators within the pocket activates K2P channels.

**Rationalizing the modulator pocket mechanism**

A natural question that arises is: How can channel activation occur via the modulator pocket? In general, functional and structural studies provide an average readout of the effect of different stimuli on the pocket, but do not reach sufficient atomic resolution to explain the dynamics of the activation event. Computational simulations are able to complement this description, allowing us to obtain atomistic details of K2P channel activation under ion permeation conditions, which is ultimately the main biological function of the channel. For example, *Lolicato et al.* [39] complemented experimental results on the suppression of C-type inactivation in TREK1 due to ML335 ligand with MD simulations, finding reduced fluctuations in the loops surrounding the ligand. Recently, using molecular dynamics, our group further studied the interactions of both ML335 and ML402 with TREK1, taking into account previous hypotheses about how putatively the activation signal is transduced to the SF.

Our simulations revealed that, in addition to the prevention of C-type inactivation (described previously by *Lolicato et al.* [39]), the ligands exert an additional effect, which is linked to a parallel mechanism that inactivates the channel: The flipping of the carbonyl forming the third binding site in the filter (S3), that induces an energy barrier high enough to prevent ion permeation through the channel (Figure 3). This flipping is precisely suppressed when the ligands are in the modulator pocket, thereby reducing inactivation events due to this mechanism. This process is directly correlated with the dynamics of the side chain of the T141 residue found at the base of the filter and interacting with the ligands: When the hydroxyl of said residue points vertically, it is able to stabilize the canonical I143 residue conformation that forms S3, which leads to a decreasing of the population of S3 flips and thus the inactive states due to such mechanism. Other molecular events linked to this mechanism are the increasing of hydration levels in the modulator pocket, and the presence of two water molecules (one in S3 and another one behind it) that help to stabilize this flipped state [64]. A similar mechanism involving carbonyl flipping had previously been described to explain mechanical gating in TREK channels through MD simulations [65,66]. In an interesting development, building on this previous evidence and our own simulations, the carbonyl-flipped state in S3 was experimentally confirmed for the prokaryotic channel KcsA using 2D infrared spectroscopy [67]. Additionally, electrophysiology experiments with sucrose gradients have confirmed the importance of water molecules at the S3 binding site in TREK channels, which is related to their inactivation process [41].

All these aforementioned results suggest that the region surrounding T141 is key to the activation of TREK1 and potentially other channels: A sequence comparison between the different subfamilies of K2P channels reveals that this lower region in the modulator pocket is conserved to be highly hydrophobic, indicating that stabilizing the residues that form it with hydrophobic and bulky motifs (given, for example, by a small molecule or a lipid) will potentially enhance the activation signal (Figure 1. Lower panel). Therefore, efforts to develop molecules that fulfill these characteristics correspond to an open research area for the modulation of these $K^+$ channels.

The upper modulator pocket has shown to play a key role in the C-type inactivation of $K^+$ channels. An example is the N147-F134 interacting pair, which consists of residues located at the top of the SF and in the P1 helix, respectively (Figure 3). The equivalent pair in KcsA corresponds to D80-W67, which is essential for the inactivation process of this prokaryotic channel [68]. Inactivated structures of *Shaker* D447-W434F and Kv1.2 with the D379-W366F mutation also suggest a coupling between filter distortion at the top and the side chain of the equivalent residue in the P1 helix. Our results indicate that both ML335 and ML402 can prevent these distortions by stabilizing the N147-F134 interacting pair in TREK1. This finding aligns with previous work by *Lolicato et al.* [39], who used X-ray crystallography and molecular dynamics to explore similar stabilization mechanisms. Since ML402 is a shorter molecule than ML335, due to the presence of a thiophene group on its upper ring (compared to the sulfonamide group in ML335), its stabilizing effect on the interacting pair is somewhat weaker, making it slightly less efficient in preventing the C-type inactivation [64].

Even more intriguing is the diversity of these equivalent interacting pairs among different K2P channels, suggesting that each channel might possess a distinct C-type inactivation mechanism. This variation is seen even within the same channel. For instance, in TREK1, the N147-F134 interacting pair is located in the modulator pocket near SF1, whereas SF2 features a D256-Y273 pair. The difference in the functional groups of these interacting side chains implies that distortions in SF2 are likely to be minor, as a hydrogen bond forms, contrasting with the weaker interaction between an amide group and a phenyl ring (similar to the substitution of a hydrogen bond donor for a phenyl ring in the prone-to-inactivation fast *Shaker* mutant D447-W434F or D379-W366F in Kv1.2).

Other critical interacting pairs are found at the residues forming the first binding site in the filter (S1) and the residue opposite it in the P2 helix. Depending on the channel, these residues may interact through hydrophobic interactions, as seen in TREK1

(F145-F271) and hERG (F627-F617), or via hydrogen bonds, as described in TREK2 (Y175-T278). In the case of TREK2, *Türkaydin et al.* [62] demonstrated that this interacting pair transduces activation signals from a distant cytosolic sensor domain to the filter [62], while *Lau et al.* [69] highlights the importance of the hydrophobic interactions in the F627-F617 displayed in the non-K2P channel hERG [69]. In a fascinating twist, these previously explained interacting pairs have been identified as key elements in the slow inactivation of Kv channels, as demonstrated by electrophysiological assays and mutations that disrupt hydrogen-bond interactions between them [70]. Given this evidence -and the parallels between C-type inactivation in Kv and K2P channels- it is plausible that a similar, though not identical, mechanism may be at play in K2Ps. This could involve alternative interactions, such as hydrophobic contacts, particularly when side-chain substitutions replace hydroxyl groups with aromatic rings, such as phenyl groups [70]. We emphasize that this structural motif has been extensively studied in the context of $K^+$ channel inactivation and has given rise to multiple hypotheses explaining the sequence variability observed in SFs. Moreover, these interacting pairs may represent promising therapeutic targets in K2P channels and deserve further investigation.

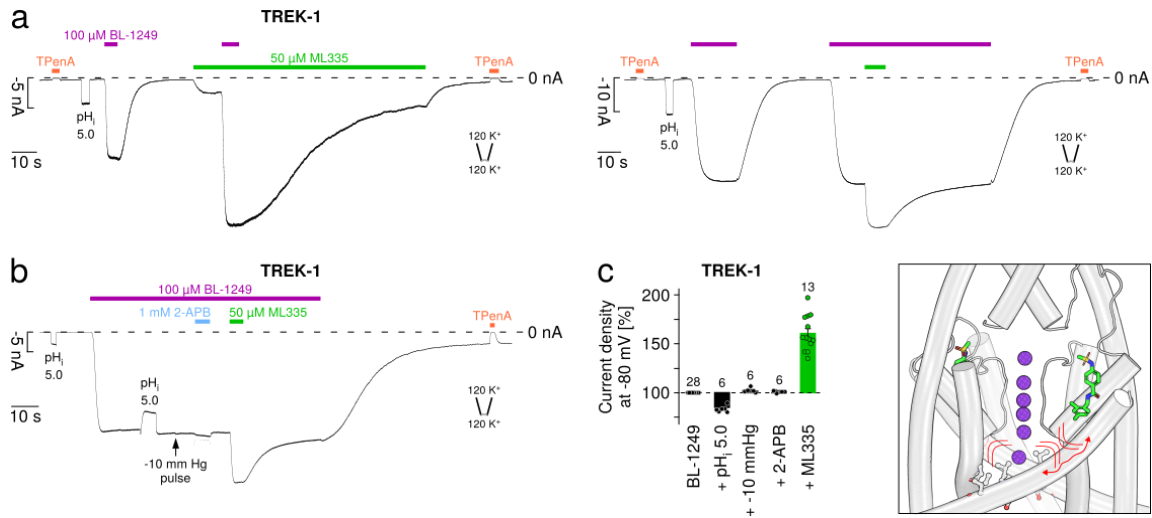

**Figure 4.** *Potential synergy between the modulator pocket and the fenestration site.* Panels a–c: Current response of the TREK1 channel upon exposure to different stimuli (pH, pressure, blockers, and activators). It is observed that ML335 (an activator binding to the modulator pocket) further increases the channel response, even when the channel is already stimulated by the NCA BL-1249. The snapshot at the bottom left next to panel c shows the crystallographic structure of TREK1, with ML335 highlighted in green and the residues interacting with BL-1249 shown as grey sticks. The red arrow indicates the region of potential crosstalk between the two sites.

An interesting feature of the modulator pocket is its potential coupling to other regulatory sites on K2P channels. For example, the fenestration site, where the inhibitor norfluoxetine (NFx) and the NCAs like BL-1249 bind [8,59], might couple to the lower region of the modulator pocket, this being an interface between both sites formed by the hydrophobic patch highlighted above as a star in Figure 1. Such coupling could, for example, enhance activation effects at both sites in a synergetic manner: a molecule in the modulator pocket might stabilize the sidechains of the lower hydrophobic patch, which would transduce such stabilization into the upper part of the fenestration site where a second agonist is bound, and vice versa (Figure 4). Moreover, it is possible that activation at the fenestration site increases the occupancy of potassium ions in the filter, then reducing the probability of water molecules occupying S3 and S0, which in turn helps an agonist bound in the modulator pocket to stabilize the canonical-conductive filter conformation -that is, with no distortions in S0 and S1 and no carbonyl flipping at S3. Both modulator pocket and fenestration site might help to escape the indirect up/down gating mechanism, as any molecule binding to the fenestration site would require M4 to adopt a down position; yet even in this state, the channel remains activated. Likewise, binding of an activator (e.g., ML335) in the modulator pocket brings the channel to the up-state, thus reducing NFx inhibition [8] . Multiple other sites have been described that can regulate K2P channels and are therefore potential coupling partners for the modulator pocket. The reader is referred to a previous review by *Natale, Deal & Minor Jr.* [71] that provides an in-depth discussion of pharmacological sites in K2P channels.

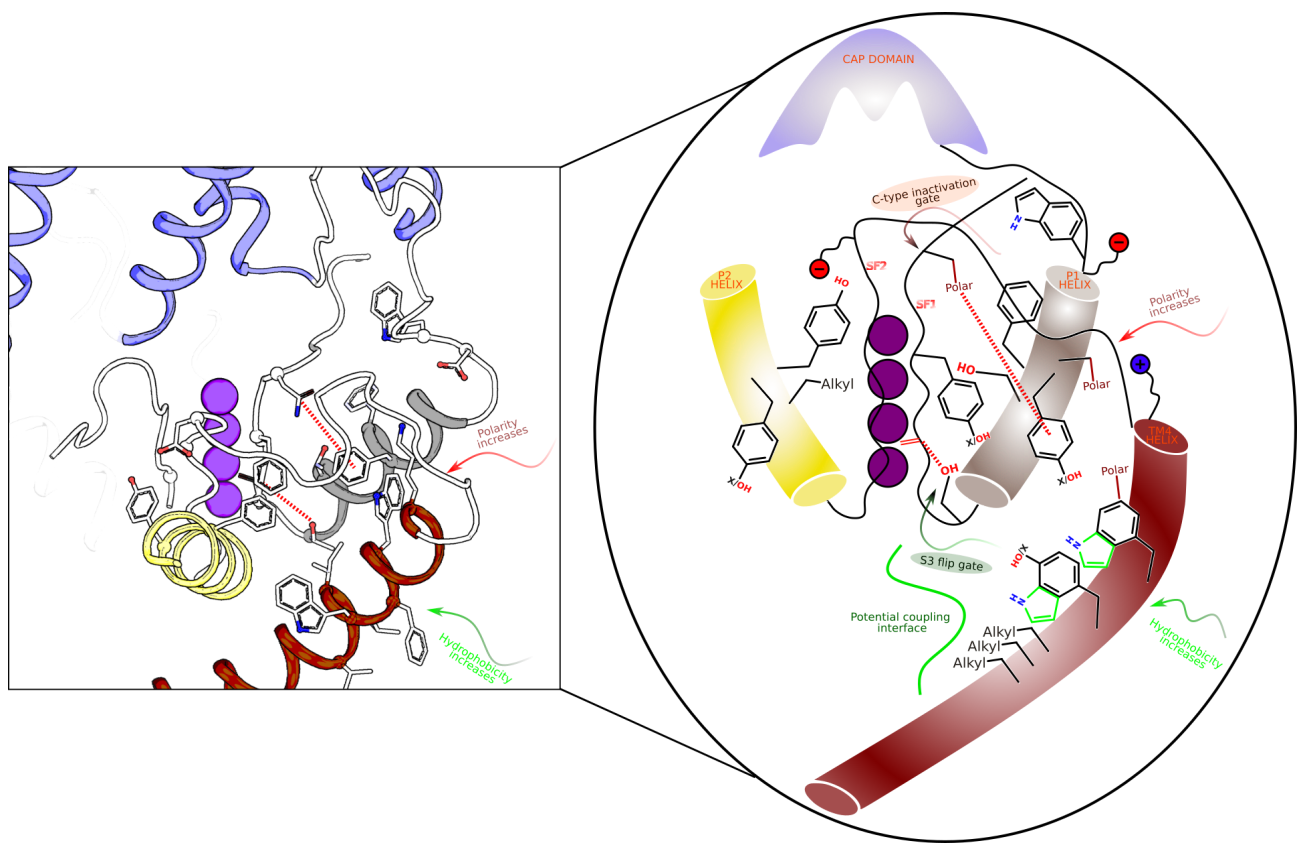

**Figure 5.** *Architecture of the modulator pocket.* **Left square panel.** Cartoon representation of the modulator pocket as shown in the crystallographic structure of TREK1. **Right oval panel.** Schematic representing, on average, the common physicochemical characteristics of the modulator pocket across different K2P channels. X substitutions indicate that the residue may lack a hydroxyl group on its phenyl ring. Green pyrrole rings indicate that the residue may be present as TRP rather than PHE or TYR. Dotted red lines represent key interactions involved in the gating mechanisms of C-type inactivation and carbonyl flipping at S3, which are potentially modulated by agonists binding to this pocket. For exceptions to this structure-binding-site-relationship, refer to the main text.

Finally, although the modulator pocket appears to be a common modulation site across all K2P channels, it is possible to find differences that likely impact how such modulation is transduced to the filter (Figure 5). For example, THIK1 and TWIK1 display a serine instead of a threonine at the base of the filter (residue T141 in TREK1), which makes the mechanism of inactivation by S3 carbonyl flipping likely to be different, as the side chain is much shorter. Regarding the P1 residue partner in the N147-F134 interacting pair -which is more or less conserved as an aromatic residue across K2P channels- TASK2 stands out as an exception, displaying an aliphatic isoleucine residue, which would likely not only affect the modulation of C-type inactivation at the top of the filter, but also allow bulkier molecules to be accommodated within the modulator pocket.

Moreover, in M4, the conservation of a positively-charged residue (K271 in TREK1) at the extreme of the helix -where it connects to the SF2-M4 loop- offers a key interaction for ML335 and ML402 modulators via cation-$\pi$ interactions (and also potentially for anionic lipid molecules). Nevertheless, notable exceptions are the TRAAK channel, which replaces the charged-residue with a polar glutamine (a sequence change that has been widely shown to explain why ML335 and ML402 modulators do not activate the TRAAK channel); and the TASK channel, which completely abolishes this positively-charged residue with a valine, again suggesting that the mechanism by which the modulator pocket activates these two channels may require molecules with different physicochemical properties that do not depend on this electrostatic interaction for binding. Furthermore, a bulky aromatic residue located in M4 -W275 in TREK1, which belongs to the modulator pocket- is exchanged by an acidic and a polar residue in the TASK2, and TWIK1 channels, respectively. These changes suggest that for those K2P channels, an amphipathic molecule capable of satisfying such polar interactions -while also partitioning into a pocket with a lateral interface to the hydrophobic membrane- is likely required. Further, comparing the hydrophobic patch at the bottom of the modulator pocket, reveals that it is more or less conserved across all channels, with some more polar substitutions in the TASK1 (T instead of L/V) and THIK1 (M instead of L/V) channels. All of the aforementioned differences between K2P channels in their modulator pocket offer an opportunity to design molecules targeting specific channels. This is important, as specificity is likely desirable for therapeutic purposes.

**Conclusions**

Since their discovery, K2P channels have proven to be important potential therapeutic targets for the treatment of cellular excitability disorders. Their unique pseudo-tetrameric topology with an extracellular domain provides versatility in the assembly of the functional channel; and the variability in the sequences that form the selectivity filter and the pore helices surrounding such filter offers a rich landscape of potential gates that -although they likely share a common core mechanism- must exhibit subtle differences depending on the channel subfamily. This is not to mention potential physically closed, but not inactivated, states of these channels, akin to the helix bundle crossing observed in Kv channels.

Stabilization of the filter is one of these key gating mechanisms of K2P channels, and its modulation has attracted considerable interest. Joint efforts have shown that binding sites for both inhibitors and activators are located around the selectivity filter, with small molecules offering an effective means of controlling its conformation.

## Acknowledgments

Edward Mendez-Otalvaro gratefully acknowledges all the researchers who, directly or indirectly, conducted the experimental and computational work cited in this review, without whom its completion would not have been possible.

## Disclosure statement

The authors declare no competing interest.

## Fundingd

This work was supported by the Max Planck Society (to E.M.O.). M.S. is funded by the Leibniz Collaborative Excellence Programme (K622/2024).

## Author contributions

E.M.O. performed the initial literature search and visualization. E.M.O., W.K., M.S. & B.L.d.G. reviewed, edited and wrote the text. E.M.O., W.K., M.S. & B.L.d.G. reviewed the visualization and E.M.O. edited the visualization. All authors have read and agreed to the published version of the manuscript.

## Data availability statement

Data sharing is not applicable to this article as no new data were created in this study.